\documentclass[12pt]{article}

\usepackage[latin1]{inputenc}

\usepackage[english]{babel}

\usepackage{epsfig}

\def\s{\sigma}

\def\a{\alpha}

\def\c{\gamma}
\def\b{\beta}
\def\d{\delta}
\def\eps{\epsilon}

\def\.{\cdot}

\def\+{\bigoplus}

\def\({\left(}
\def\){\right)}
\def\[{\left[}
\def\]{\right]}
\def\l.{\left.}
\def\r.{\right.}

\def\<{\left\langle}
\def\r|{\right|}
\def\>{\right\rangle}
\def\l|{\left|}

\def\beq{\begin{equation}}
\def\eeq{\end{equation}}
\def\bea{\begin{eqnarray}}
\def\eea{\end{eqnarray}}
\def\nn{\nonumber \\ &&}

\def\ber{\begin{array}}
\def\eer{\end{array}}

\newcommand{\psl}{ \mkern-6mu \not \mkern-3mu p}
\newcommand{\ksl}{ \mkern-6mu \not \mkern-3mu k}
\newcommand{\qsl}{ \mkern-6mu \not \mkern-3mu q}

\begin{document}

\title{An off-shell I.R. regularization strategy in the analysis of collinear divergences}
\author{Carlo M. Becchi and Alessandra Repetto\\
Dipartimento di Fisica, Universit\`a di Genova,\\
Istituto Nazionale di Fisica Nucleare, Sezione di Genova,\\
via Dodecaneso 33, 16146 Genova (Italy)}
\maketitle
\abstract{
We present a method for the analysis of  singularities of Feynman amplitudes based on the Speer sector decomposition of the Schwinger parametric integrals combined with the Mellin-Barnes transform. The sector decomposition method is described in some details.

We suggest the idea of applying the method to the analysis of collinear singularities in inclusive QCD cross sections in the mass-less limit regularizing the forward amplitudes by an off-shell choice of the initial particle momenta.

It is shown how the suggested strategy works in the well known case of the one loop corrections to Deep Inelastic Scattering.}

\section{Introduction}

The success of Feynman's parton model \cite{Feynman:1969} has driven  the search for a field theory formulation of the parton hypothesis. The discovery of asymptotic freedom  \cite{Gross:1973ju}  of unbroken non-abelian gauge theories has given a clear indication of the field content of the Feynman model. Partons are fermions, quarks, and vector bosons interacting through exchanges of vector bosons, gluons.

However, partons, therefore quarks and gluons, should be bound in a hadronic initial state. Therefore a high energy partonic process should be characterized by two scales, the high energy (momentum transfer) scale and the binding energy scale, which should be less than few hundred MeV.

If one tries to exploit perturbative field theory in order to compute corrections to the naive free parton model, one finds typically logarithmic factors in the ratio of the two scales, which is about one thousand. These logarithmic factors appear as multipliers of the strong structure constant $\a_s$. Thus the validity of perturbation theory becomes less obvious in much the same way as the idea of parton independence, that is factorization.
The only "simplification" induced by the presence of the two above mentioned scales is that at least for the quarks up and down, whose mass is few MeV, one can consider quarks as massless as gluons are. Therefore the field model should be considered in first (free) approximation scale invariant and the radiative corrections should violate Bjorken scale invariance due to renormalization. This is indeed what happens and the corrections to Bjorken scaling \cite{Bjorken:1968dy} give a strong evidence in favour of the non-abelian QCD model. 

However the calculations of radiative corrections together with the presence of the above mentioned scales, one of which is associated with the process, the other appearing in the hadronic wave function, have posed non-trivial technical difficulties which have been overcome thanks to the discovery of dimensional regularization \cite{'tHooft:1972fi}~\cite{BM} and of the corresponding minimal renormalization schemes. The simplicity of computation is only one of the advantages which have strongly favoured the use of the dimensional method. A second, very important, advantage is that it doesn't break gauge invariance and applies without problems to the, however singular, quark and gluon mass-shell amplitudes. This is very important in view of parton factorization hypothesis. The dimensional method is consistent with the description of the hadron as a gas of free quarks and gluons whose distribution function has to be computed on the basis of consistency conditions of different scale choices.

The QCD parton amplitudes are however infrared (IR) singular,  this happens in particular, in the study of deep inelastic scattering (DIS).  For this reason a huge amount of work has been done aiming at the analysis of the singularities of Feynman diagrams as functions of the kinetic invariants, that is, of the invariant squared partial sum of the momenta carried by the external vertices. The study of collinear singularities gives information through the Altarelli-Parisi \cite{Altarelli:1977zs} approach on the parton distribution function of e.g. a target baryon.
Infrared singularities appear in a dimensional regularized   mass-shell massless parton amplitude  as poles in the origin of the  complex plane of the dimensional regularization parameter $\eps=\frac{d-4}{2}$. Thus in the dimensional scheme IR singularities appear  in much the same way as UV singularities do. This implies that one has to avoid IR-UV singularity mixings by first subtracting UV divergences. 
 One should however keep firmly in his mind that, while the  UV singularities are renormalized by a redefinition of few parameters, IR ones are associated with a bad choice of the initial and final states of the process. 
 
An important result concerning the Feynman graph singularities is that in the Schwinger parametric representation \cite{Speer:1970ss}~\cite{Smirnov:2008} the singular parts of a diagram are confined in particular sectors of the parametric space. 
This induces a remarkable simplification in the singularity analysis but until recently it has not been systematically exploited in calculations.(See however \
\cite{binoth} \cite{Carter:2010hi} \cite{Smirnov:2009pb} and references therein).  

Beyond purely technical aspects, an easy tool for the computation of the infrared singularities of the Feynman amplitudes allows a better physical insight into Parton Physics. From the point of view of the physical interpretation, one should take into account that partons are identified with bound, and hence off-shell quarks and gluons. Notice that considering off-shell amplitudes one introduces a kind of IR regularization since the infrared singularities are regularized by a suitable off-shell (Euclidean) choice of external momenta and are therefore well separated from the (dimensionally regularized) UV singularities. 
Hence the idea is to regularize the parton amplitudes choosing off-shell initial momenta \cite{Carlitz:1988ab}~\cite{Dorn:2008dz}. The singularities of the amplitudes when the parton momenta go on-shell are related to the Altarelli-Parisi splitting functions. 

 However considering single parton amplitudes with off-shell initial particle states presents a further difficulty. Indeed it is fairly well known that off-shell charged particle amplitudes are not "gauge invariant".  As a matter of fact, strictly speaking the single parton contributions are not physically meaningful and hence parton factorization, which is the basis of the parton model, should be taken "cum grano salis". This difficulty is overcome by the fact that, contrary to  generic off-shell amplitudes, their mass-shell singular and finite parts are gauge invariant and independent.

We hope that applying the present idea to a physical situation, such as the first order corrections to Deep Inelastic Scattering, will help us to make this working hypothesis clear.

In this work we present the off-shell analysis in the light of few, more or less, recent progresses in Feynman graph computations. These are essentially based on the extended use of Mellin-Barnes transform \cite{M-B}, aiming at the singularity analysis, and the sector decomposition of parametric Feynman integrals, mentioned above.

A further relevant aspect of our analysis lies in the evaluation of the coefficients of collinear singularities for Euclidean values of the kinetic invariants. These coefficients, which are analytic functions of the kinetic invariants, are analytically continued to the physical region where their limiting values automatically give well defined distributions, in our case, for the Altarelli-Parisi splitting functions. No problem appears on the border of the physical region where it is often spoken of soft singularities. We think that the presented example clearly exhibits this fact which is however completely general.

Notice that a generic not renormalized amplitude might present UV singularities which might combine with mass singularities. The presence of UV and mass singular contributions might appear in the case of multiloop diagrams with UV divergent subdiagrams. However, once  these UV divergences are minimally subtracted in the Breitenlohner-Maison \cite{Breitenlohner} scheme they should not interfere with the mass singularity analysis and, in particular, they do not interfere in the one-loop case.

\section{Feynman amplitudes, Schwinger parametric form and Speer-Smirnov sectors}
In this section we give a short description of Speer-Smirnov's sectors.
We start considering a generic connected \footnote{Notice that a diagram is a set of lines {\bf and} vertices, thus a connected diagram must contain, for any pair of its points, enough lines and {\bf vertices} to form a continuous path joining them.} Feynman amplitude in $d$ space-time dimensions associated with a diagram with $I$ lines and $L$ loops (and $V=I-L+1$ vertices). If  the amplitude corresponds in the momentum  representation to  a $Ld$ dimensional integral of the product of a numerator $N$ with momentum dimension $d_N$ and $I$ scalar propagators,  in the  Schwinger parametric form the same amplitude is \cite{Speer:1970ss} \cite{BM}:
\beq
\tilde{A_G}(p)=\frac{i^{I+L-Ld}}{(4\pi)^{Ld/2}}\,C_V\sum_{\tau}\sum_{a=0}^{\[\frac{d_N}{2}\]} \sum_{b=0}^{d_N-2a}\Theta_{a,b,\tau}(p) \,I_{G,(a,b,\tau)}(p),
\label{princ2}
\eeq
where $C_V$ is a suitable coefficient depending on the coupling constants and combinatorial factors,   the sum over $\tau$ accounts for the different irreducible tensors contributing to $\tilde{A_G}(p)$ ,  $\Theta_{a,b,\tau}$ is a homogeneous polynomial  of degree \beq d_{\Theta}=d_N-2a\nonumber\eeq in the components of the momenta $p$ entering into the diagram through its external vertices. We denote by $[X]$ is the integer part of the  positive real number $X$. The parametric integral factors are:
\beq
I_{G,(a,b,\tau)}(p)= \mu^{2L\eps}\Gamma(I-\frac{Ld}{2}-a ) \int \frac{d\mu(\b)\,\prod_{i=1}^I \b_i^{\lambda_{i,\tau}}}{P_G(\b)^{\frac{d}{2}(L+1)+b-I}D_G(\b,p)^{I-\frac{Ld}{2}-a }}
\label{Ib}
\eeq
where
\begin{itemize}
\item
the integration domain is  identified with the quotient space of the positive sector of the $I$-dimensional Cartesian space and a positive semi-line, that is, with the positive sector of the $I-1$ - dimensional projective space. Therefore, for example, one can write:
\beq d\mu(\b)=\(\prod_{i=1}^I d\b_i\)\delta(1-\sum_{i=1}^I\b_i),\label{cubic}\eeq however this is not the choice we shall use in the following,
\item $P_G(\b)$ is a homogeneous polynomial  of degree $L$ in the $\b$ variables,
\item $D_G(\b,p)$ is a quadratic form in the  momenta entering the external  vertices $p$, whose coefficients are homogeneous polynomials of degree $L+1$ in  $\b$,
\item the exponents $\lambda_{i,\tau}$ are integer non negative numbers and satisfy the sum rule:$\ 
\sum_i^I\lambda_{i,\tau}=L(b-a)-a\ .$
\end{itemize}
The construction rules of the integrand, in particular, of the Symanzik functions $P_G(\b)$ and $D_G(\b,p)$  together with the determination of  the exponents $\lambda_{i,\tau}$  can be found e.g. in \cite{Speer:1970ss} and \cite{BM} where one also finds the subtraction rules for the UV divergences. In \cite{repetto} one finds the details and proofs of the ensuing analysis.

A very general IR power counting theorem states that: {\it   in a mass-less theory, such as QCD, in which  fields have positive mass dimensions and  internal vertices have dimension four, an amplitude with space-like external momenta has no infrared divergences  if the external momenta are non-exceptional, i.e. no partial sum vanishes.} \cite{Zimmermann} We shall call a theory of this kind, to which the present paper is devoted, {\it IR safe mass-less theory.}

 Now we come to the sector decomposition of the integration domain and to the choice of the parameters $\b$ in each sector \cite{Smirnov:2008}. This is strictly related to the structure of the diagram. We shall try to give an idea of the construction concentrating  on one-particle irreducible (1P-I) diagrams. 
 
 Some basic definitions are in order. We call {\it irreducible} a connected diagram which
cannot be broken  into  two connected components deleting a line or a vertex, we also consider irreducible the {\it trivial} case of a diagram consisting in a single line. A connected reducible diagram is naturally reduced to a collection of {\it parts} ({\it pieces} \cite{Speer:1970ss}) which are irreducible. Given an irreducible diagram $G$ we call {\it link} a connected sub-diagram $\lambda_G$ of diagram  $G$ which contains all the external vertices of $G$ and is minimal, in the sense that it does not contain anymore all the external vertices if one omits one of its parts. 

The Speer-Smirnov construction is based on the identification of all the {\it singularity families} of $G$ (s-families) which are maximal sets of $G$-sub-diagrams which are, either links, or irreducible and which do not overlap, that is, either they have no lines in common, or are contained into one another. In particular the links are always contained into one another. The whole diagram $G$ is the first element of any s-family of $G$.

It turns out that a s-family is identified assigning a set $C$ of lines of $G$,  that we call  {\it complete    cut},  whose deletion breaks $G$ into two tree sub-diagrams each of which contains at least one external vertex. The first step in the construction is the choice of an order in $C$. Then the elements of the s-family are identified deleting the lines belonging to $C$ in the chosen order\footnote{In the existing literature instead of cuts one considers the complementary set of lines which are called 2-trees. }. The first deletion reduces $G$ into the union of a link  and a set of  irreducible sub-diagrams,  the other parts of what remains of $G$ after the line deletion. Then one continues reducing links into sub-links and further  irreducible sub-diagrams. When the last link is broken into two components, each containing one or more external vertices and these components are decomposed into their irreducible parts, one is left with a set of trivial and non-trivial irreducible sub-diagrams. Now one starts deleting  further lines of the non-trivial elements of the set,  breaking all their loops.  After the last line of $C$ has been deleted, one is left with a set of tree sub-diagrams whose parts are trivially irreducible. The s-family is identified with the set of links and, both non-trivial, and trivial  irreducible sub-diagrams that have been generated during this sequence of deletions and reduction into parts.
\begin{figure}[ht]
\hskip 1.7cm\includegraphics*[scale={0.46} , clip=false]{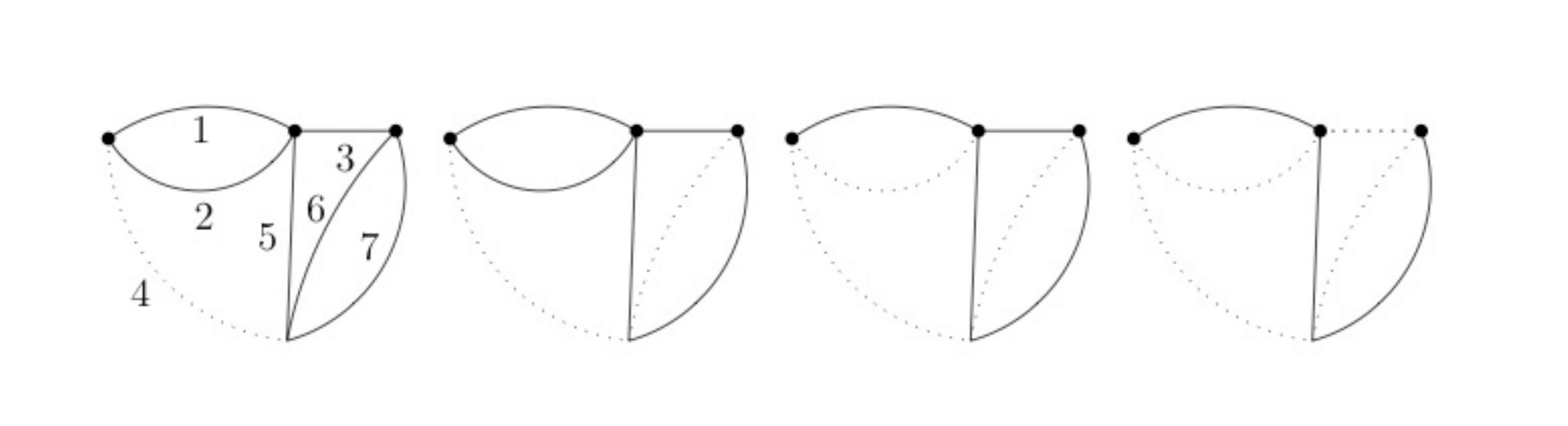}
\caption{\small An example of a s-family in an irreducible 4-loop diagram. The black points are external vertices. The solid line sub-graphs denote the links belonging to the s-family which also contains the whole diagram and  lines $1$ and $7$. The complete cut in the chosen order is the set of lines: $(4, 6, 2, 3, 5)$, the reduced cut is $(2, 3, 4, 6)$. A second example of s-family corresponds to the ordered cut $(4, 5, 3, 2, 6)$ and reduced cut $(2, 4, 5, 6)$. The reader can easily draw the corresponding s-family which now contains 2 instead of 4 links, not considering the whole diagram.}
\end{figure}

The reader can convince himself that,
\begin{itemize}
\item a s-family $F$ in $G$ contains $I$ elements, if $I$ is the number of  lines in $G$,
\item the number of lines in a complete cut is equal to $L+1$ if, as above, $L$ is the number of loops,
\item one can establish a one-to-one correspondence $l_\c$  between the elements of $\c\in F$ and the lines of $G$. This correspondence is obvious for the trivial elements of $F$. In the case of a non-trivial element $\c\ ,$ $l_\c$ is identified with the first line of the ordered complete cut belonging to $\c$.
\end{itemize}
Assigning to every element $\c\in F$ a variable $t_\c$ with $0\leq t_\c\leq 1$ and  setting $t_G=1$ and $\b_l\equiv\prod_{  l\in \c\in F }t_\c\ ,$ one defines the
 Speer-Smirnov parametrization of a sector of the $I-1$ dimensional integration domain in Eq.(\ref{Ib}). This is apparently different from that appearing in Eq.(\ref{cubic}), indeed for the first line of the cut one has $\b_{1}=1$ and hence $\sum_l\b_l\geq 1\ .$
 Notice that with the new choice of variables each integration domain corresponds to a $I-1$ dimensional hypercube in the $t_\c$ space.  The reader can verify that the sectors corresponding to different s-families are disjoint and the set of sectors covers the whole integration domain.
 
 On the basis of the above construction of s-families and sectors the parametric integral appearing in Eq.(\ref{Ib}) decomposes into the sum over sectors, and hence s-families $F$:
 \beq
I_{G,(a,b,\tau)}(p)=\sum_F I^F_{G,(a,b,\tau)}(p)
\eeq
with
\beq I^{(F)}_{G,(a,b,\tau)}(p)=\mu^{2L\eps}\,\Gamma(I-\frac{Ld}{2}-a)\int \frac{d\mu^{(F)}_G\,\prod_{i=1}^I \b_i^{\lambda_{i,\tau}}}{P_F^{\frac{d(L+1)}{2}-I+b}(\b)D_F^{I-\frac{Ld}{2}-a}(p, \b)}.
\label{I_Sb}
\eeq
In the last expression the measure $d\mu^{(F)}_G$ and the Symanzik functions $P_F$ and $D_F$ depend on $F$, which is the s-family identifying the particular sector.
The  integration measure of the sector amplitude,  is 
\beq
d\mu^{(F)}_G=\prod_{l=2}^I d\b_{l}=\prod_{\c\in{ F}\ ,\c\neq G}t_{\c}^{I(\c)-1}\,dt_{\c},
\label{mu}
\eeq
where the product runs on all the elements of the s-family, except for the diagram $G$ itself,  and $I(\c)$ is the number of lines contained in the sub-diagram $\c$.

In order to establish the structure of the Symanzik function $P_F$ we have to introduce a further kind of sub-sets of lines of $G$ that we call {\it reduced cuts} and label by $\hat C$. A reduced cut is  the complement of a maximal tree sub-diagram of $G$, it is a set of $L$ lines whose deletion reduces $G$ to a tree diagram. Once identified all the reduced cuts, one has:
\beq
P_F(t_\c)=\sum_{\hat{C}}\prod_{\c\in F}t_{\c}^{I(\c\cap \hat{C})}\ ,\label{Ps}\eeq
where  $I(\c\cap \hat{C})$ is the number of the lines in the intersection of the element $\c$ of the s-family  and the reduced cut $\hat{C}$.

Let us now  consider $D_F$.  On account of the external momentum flow through the diagram, given a complete cut $C$, which does not necessarily coincide with $C_F$, one defines, up to a sign, the external momentum $p_C$ crossing  the complete cut. Then one has:
\beq
D_{F}(p, t_\c)=\sum_{C }p^2_C \prod_{\c\in F}t_{\c}^{I(\c\cap {C})}\ ,\label{Ds}\eeq
where
$I(\c\cap C)$ is the number of the lines in the intersection of the element of the s-family $\c$ and the complete cut $C$.

Now we discuss the behavior of the Symanzik functions in a given sector $F$.  Since both $P_F$ and $D_F$ are polynomials in the $t_\c$ variables and since $0\leq t_\c\leq 1\ ,$ it is possible to identify, given the s-family $F$, the larger monomial in $P_F$ and the larger coefficients of the kinetic invariants in $D_F$. 

We consider first $P_F$ given in Eq.(\ref{Ps}) as a sum over  the reduced cuts and  we single out the dominant contribution. This corresponds  to the reduced cut $\hat C_F$ which  is obtained from the ordered complete cut $C_F$ identifying the s-family $F$ omitting the line whose deletion breaks the last link. This 
line belongs to all the links contained in $F$ and  gives a maximal tree sub-diagram when added to the 2-tree diagram made of the lines of $G$ not belonging to $C_F$; this is not true for the ensuing lines in the ordered cut.
The monomial corresponding to $\hat C_F$ in $P_F$ is given by $\prod_{\c\in F}t_{\c}^{I(\c\cap\hat C_F)}$. This is larger  than the other monomials inside  the hypercubic sector. Indeed we have:
\beq P_F(t_\c)=\prod_{\c\in F}t_{\c}^{I(\c\cap\hat C_F)}(1+\sum_{\hat C\not=\hat C_F}\prod_{\c\in F}t_\c^{\d_{\hat  C, F}(\c)})\ ,\label{Pss}\eeq with $\d_{\hat  C, F}(\c)\equiv I(\c\cap \hat C)-I(\c\cap \hat C_F)\geq 0$ and $\sum_{\c\in F}\d_{\hat  C, F}(\c)>0\ .$

More important for the IR singularities is the corresponding decomposition of $D_F$. In order to analyze the behavior of this function in the sector corresponding to the s-family $F$, let us recall, first of all, that one has to study separately the coefficients of the various kinetic invariants. Since $F$ corresponds to a given  complete cut $C_F$ with a chosen order,  the corresponding kinetic invariant $p_{C_F}^2$ naturally  plays a particular role in our analysis. However its coefficient is  in general the sum of several monomials, indeed,  same momentum $p_{C_F}$ crosses several complete cuts. If we denote by ${\cal C}_F$ the set of these cuts, each of which gives a contribution equal to $\prod_{\c\in F}t_{\c}^{I(\c\cap C)}$ to the coefficient  of the same kinetic invariant $p_{C_F}^2$, one can show that in the sector $F$ the contribution from the cut $C_F$ is dominant in the sense that:
\bea D_F(p, t_\c)&& =\prod_{\c\in F}t_{\c}^{I(\c\cap C_F)}\[p_{C_F}^2(1+\sum_{C\in {\cal C_F}, C\not=C_F}\prod_{\c\in F}t_\c^{\d_{C, F}(\c)})\right.\\ && \nonumber\left.+\sum_{C\not\in {\cal C_F}}p_C^2\prod_{\c\in F}t_{\c}^{\d_{C, F}(\c)}\]\equiv \prod_{\c\in F}t_{\c}^{I(\c\cap C_F)}\[p_{C_F}^2N_F(t_\c)+M_F(p,t_\c)\]\ ,\label{Dss}
\eea
where  $\d_{ C, F}(\c)\equiv I(\c\cap C)-I(\c\cap C_F)\geq 0$ and $\sum_{\c\in F}\d_{ C, F}(\c)>0\ .$

Combining  our results together we have for the un-subtracted sector amplitude introduced in Eq.(\ref{I_Sb}) the following expression:
\beq I^{(F)}_{G,(a,b,\tau)}(p)=\mu^{2L\eps}\,\int{\Gamma(I-\frac{Ld}{2}-a) \ d\bar\mu^{(F,d)}_{G,(a,b,\tau)}\over \[p_{C_F}^2N_F(t_\c)+M_F(p,t_\c)\]^{I-\frac{Ld}{2}-a}}\ ,
\label{I_St}
\eeq where:
\beq d\bar\mu^{(F,d)}_{G,(a,b,\tau)}={\prod_{\c\in{ F}\ ,\c\neq G}t_{\c}^{E^{(F,d)}_{G,(a,b,\tau)}(\c)-1}\,dt_{\c}\over(1+\sum_{\hat C\not=\hat C_F}\prod_{\c\in F}t_\c^{\d_{\hat  C, F}(\c)})^{\frac{d(L+1)}{2}+b-I} }\ ,
\eeq
 and 
\bea E^{(F,d)}_{G,(a,b,\tau)}(\c)&&=\sum_{l_i\in\c}\lambda_i +I(\c)+(L{d\over2}-I+a)(I(\c\cap C_F)\nn-I(\c\cap \hat C_F)) +(a-b-{d\over2})\ I(\c\cap \hat C_F)\ .\label{M_St}
\eea

Now we see that in an IR safe theory and   for Euclidean non-exceptional external momenta, that is, all $p_C^2<0$ and $d=4$, the amplitude $ I^{(F)}_{G,(a,b,\tau)}(p)$ diverges, either if $I-\frac{Ld}{2}-a\leq 0$, and hence it is a primitively UV divergent, or one or more  $E^{(F,d)}_{G,(a,b,\tau)}(\c)\leq 0$, and hence $G$ contains sub-diagrams which are  UV divergent in the sector $F$. Indeed, as mentioned above, the amplitude cannot be affected with IR divergences in the Euclidean-non-exceptional domain and in this domain the denominator in the integrand in Eq.(\ref{I_St}) does not vanish.
The above divergences must be cured using the Breitenlohner-Maison subtraction method. Applying the above formulae to the example presented in figure (1), and in particular considering the sectors corresponding to the two presented s-families, the reader can easily verify that in the first sector one has a contribution which is only quadratically and superficially UV divergent, while that from the second sector also presents UV divergences associated with the sub-diagram made of  lines $1$ and $2$ and that made of lines $6$ and $7$. Thus the subtraction procedures in the two sectors must be different. We shall not discuss this point anymore since in the examples we are going to discuss only primitive divergent terms appear which do not contribute to the IR collinear singularities we are studying. Thus in particular in our examples the measure $d\bar\mu^{(F,d)}_{G,(a,b,\tau)}$ is integrable in the  sector $F$.

In order to complete our study we have to change  our point of view. Instead of looking at a given sector  considering the IR singularities developed by the contribution of the sector to the amplitude, we must consider one, or more than one, kinetic invariants whose vanishing together characterizes the IR singularity.  We have seen that once the sector is given, this identifies a complete cut and hence a kinetic invariant coinciding with the square momentum crossing the cut, in general a kinetic invariant corresponds to more than one sector and hence several sectors contribute to the IR singularities associated with a set of vanishing kinetic invariants. 

Let us consider a set of kinetic invariants with the same (or possibly proportional) negative value $-\xi^2$ and let ${\cal F}_{\xi}$ be the set of s-families $F_\xi$, and ${\cal C}_{\xi}$  that of the corresponding complete cuts $C_{ F_\xi }$, such that the kinetic invariants $|p_{C_{ F_\xi }}^2|$  belong to the above set and hence are equal to or smaller than $\xi^2$.
With these definitions the $\xi^2\to 0^+$ IR singularity appears in the sum over the sectors $F_{\xi}$ of $I^{F_\xi}_{G,(a,b,\tau)}(p)$, that is in:\beq I^{\xi}_{G,(a,b,\tau)}(p)\equiv \sum_{F_\xi\in{\cal F}_{\xi}}I^{(F_\xi)}_{G,(a,b,\tau)}(p)\eeq with:
\beq
I^{(F_\xi)}_{G,(a,b,\tau)}(p)=\mu^{2L\eps}\,\int d\bar{\mu}_{G,(a,b,\tau)}^{F_{\xi},d}\frac{\Gamma(I-\frac{Ld}{2}-a)}{(-\xi^2\, N_{F_{\xi}}(t_{\c})+M_{F_{\xi}}(p,t_{\c}))^{I-\frac{Ld}{2}-a}},
\label{Ixi}
\eeq
where
\beq N_{F_{\xi}}(t_{\c})=1+\sum_{C_{ F_{\xi}}\neq C\in{\cal C}_{\xi}}\prod_{\c\in F_\xi}t_{\c}^{\d_{C, F_\xi}(\c)} \label{N2} \eeq
is a positive polynomial
and 
\beq
M_{F_{\xi}}(p, t_{\c})=\sum_{ C\notin {\cal C}_{\xi}}p_C^2\prod_{\c\in  F_\xi}t_{\c}^{\d_{C,  F_\xi}(\c)}
\label{M2}
\eeq
where, following our strategy, all the $p_C^2$ are taken negative. Therefore
$M_{F_\xi}$ is a negative polynomial which in general vanishes on the boundary points of the hypercube.

Our purpose is to analyze the behavior of $I^{F_\xi}_{G,(a,b,\tau)}$ when $\xi^2\rightarrow 0^+$ and this can be done employing the Mellin-Barnes transform \footnote{An analogous procedure, aiming at the analyses of power expansions of Feynman amplitudes, has been  introduced by Pilipp \cite{pilipp}.}.

Formally one can write:
\bea
 I^{F_\xi}_{G,(a,b,\tau)} &&= \frac{1}{2\pi i}\int d\bar{\mu}_{G,(a,b,\tau)}^{F_{\xi},d}\ (-M_{F_{\xi}}(p,t_{\c}))^{a+\frac{Ld}{2}-I}\nn
 \int_{{\cal P}\[-i\infty,+i\infty\]} d\s\ \Gamma(\s)\Gamma(I-\frac{Ld}{2}-a-\s)\(-\frac{\xi^2 N_{F_{\xi}}(t_{\c})}{M_{F_\xi}(p, t_{\c})}\)^{-\sigma}\ ,
\label{MB1}
\eea where the path ${\cal P}\[-i\infty,+i\infty\]$ is a continuous line going from $-i\infty$ to $i\infty$ leaving the poles of $\Gamma(\sigma)$ on its left-hand side and the poles of $\Gamma(I-\frac{Ld}{2}-a-\s)$ on its right-hand side.
This path should be closed around  the poles of $\Gamma(\s)$ if  $|\xi^2 N_{F^{\xi}}(t_{\c})|<|M_{F^{\xi}}(p,t_{\c})|$ and around  the poles of $\Gamma(I-\frac{Ld}{2}-a-\s)$ if  $|\xi^2 N_{F^{\xi}}(t_{\c})|>|M_{F^{\xi}}(p,t_{\c})|$. However, in order to profit of the Mellin-Barnes formula, we have to change the order of the integrals considering 
\bea
 I^{F_\xi}_{G,(a,b,\tau)}&&=\int_{{\cal P}\[-i\infty,+i\infty\]} \frac{d\s}{2\pi i} \,(\xi^2)^{-\s}\,\Gamma(\s)\Gamma(I-\frac{Ld}{2}-a-\s)\nn
 \int d\bar{\mu}_{G,(a,b,\tau)}^{F_{\xi},d}\ (-M_{F_{\xi}}(p, t_{\c}))^{a+\frac{Ld}{2}-I}\(-{ M_{F_{\xi}}(p, t_{\c})\over N_{F_{\xi}}(t_{\c})}\)^{\sigma} 
 \label{MB2}
\eea  with suitable condition on how  the path should be closed.
Taking into account that $N_{F_{\xi}}(t_{\c})$ is positive and bounded in the hypercube $(0\leq t_{\c}\leq 1)$, if $-M_{F_{\xi}}(p,t_{\c})$ were bounded from below by a positive number, for small enough $\xi^2$ the two formulae in Eq.(\ref{MB1}) and in Eq.(\ref{MB2}) would coincide with the path closed around the poles of $\Gamma(\s)$. This is however not always the case, since $M_{F_{\xi}}(p, t_{\c})$ might vanish on the sector boundary.

A  consequence of this vanishing is the fact that the integral
\beq
f(\s)=\int  d\bar{\mu}_{G,(a,b,\tau)}^{F_{\xi},d}\ (-M_{F_{\xi}}(p,t_{\c}))^{a+\frac{Ld}{2}-I}\(-\frac{M_{F_{\xi}}(p,t_{\c})}{N_{F_{\xi}}(t_{\c})}\)^{\sigma},
\eeq
instead of being an analytic function of $\s$, might develop a finite number of poles. 

It follows that, while in the analytic case the path in Eq.(\ref{MB2}) should be closed around the poles of $\Gamma(\s),$ in the second case the path must leave not only the poles of $\Gamma(\s)$, but also those of $f(\s)$, on its left-hand side and those of $\Gamma(I-\frac{Ld}{2}-a-\s)$ on its right-hand side and the amplitude must be computed closing the path around the poles of $\Gamma(\s)$ and those of $f(\s)$.

Let us make this more clear presenting a trivial example:
\bea&&\int_0^1\frac{1}{\xi^2+x}dx=\int_{{\cal P}\[-i\infty,+i\infty\]} \frac{d\s}{2\pi i} \,(\xi^2)^{-\s}\,\Gamma(\s)\Gamma(1-\s)\int_0^1 dx\, x^{\s-1}\nn=
\frac{1}{2\pi i}\int d\s
\frac{(\xi^2)^{-\s}}{\s}\,\Gamma(\s)\Gamma(1-\s)\nn=-\log(\xi^2)-\sum_{n=1}^{\infty}\frac{(-1)^n}{n!}\xi^{2n}
=\log\(\frac{1+\xi^2}{\xi^2}\)\ ,\nonumber
\eea
The pole from the $x$-integral transforms that of $\Gamma(\s)$ in $\s=0$ from the first to the second order and hence produces the term $-\log(\xi^2)$ which gives the singular part of the original integral. The remaining power series gives $\log(1+\xi^2)$ which is the analytic part of the integral around the origin.

\section{Collinear singularities in DIS}

We shall use the lecture notes \cite{Ellis:1991qj} as a basic reference using the same notation. In the framework of the naive parton model the hadronic tensor is computed  in terms of the parton (quark) distribution function $q_0(y)$ and of the partonic tensor $W^p_{\mu,\nu}(k,q)$ using the formula:
\beq
W_{\mu,\nu}(k,q)=\int_0^1dy\,q_0(y)\,W^p_{\mu,\nu}(y\,k,q),
\eeq
where $y$ is the fraction of proton momentum taken by the quark.

In QCD the naive partonic tensor is computed from the tree approximation forward quark-virtual-photon Compton scattering amplitude by means of:
\beq 
W^p_{\mu,\nu}(k,q)=-\frac{1}{4\pi}Im\(\frac{Tr\[\ksl\c_{\mu}(\ksl+\qsl)\c_{\nu}\]}{(k+q)^2+i\eta}\)
\eeq
Thus, according to the optical theorem, the partonic tensor is related to the absorptive part of the tree approximation forward Compton amplitude in figure (\ref{t-a}). 
{\begin{figure}[ht]\hskip 4.7cm
\includegraphics*[scale={1.5} , clip=false]{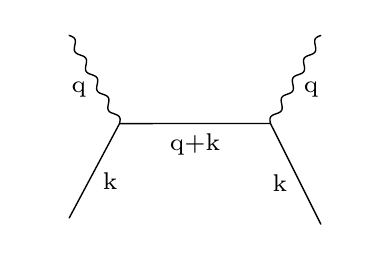}\vskip -.5 cm\caption{\small Tree diagram.} \label{t-a} \end{figure}}

Taking into account the radiative corrections, one has to consider the forward virtual photon-quark scattering amplitude corrections corresponding to the diagrams in figure (\ref{gr}).

{\begin{figure}\hskip 1.5cm
\includegraphics*[scale={1.3} , clip=false]{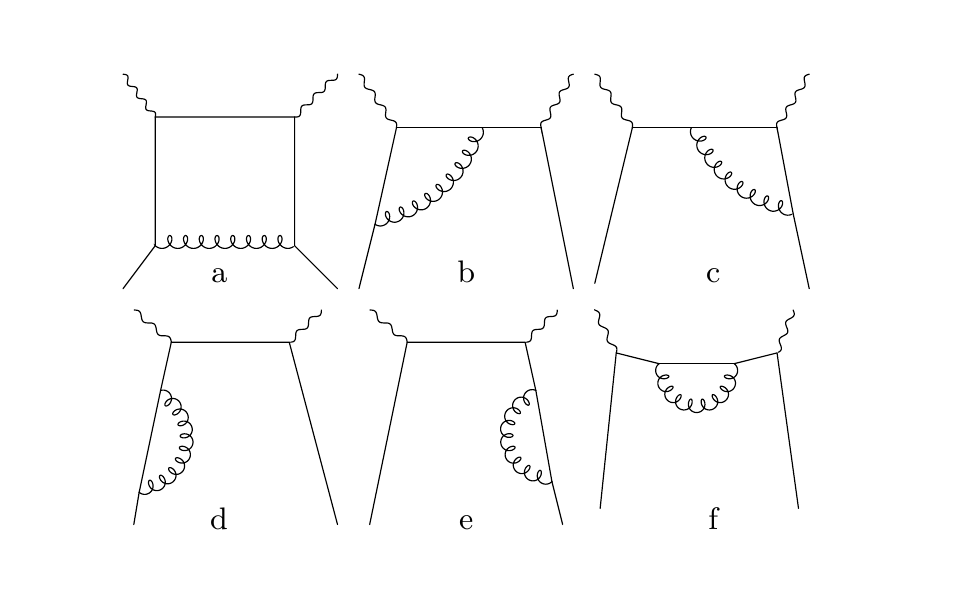}\vskip -.8 cm\caption{\small One loop corrections.} \label{gr} 
\end{figure}}
In all these diagrams the initial  and final states coincide and  consist of a quark with momentum $k$ and  a virtual photon with momentum $q$.

An elastic $2\to2$ scattering amplitude depends on six independent kinetic invariants. Four of them are the square momenta (masses) of the four external legs, that is of initial and final, possibly virtual, particles. Further invariants are the square center of mass energy, i.e. the Mandelstam variable $S$, and the square momentum transfer, i.e. the Mandelstam variable $T$.

We denote the independent kinetic invariants  by $k^2$, the square momentum of the quark, $q^2$, the square momentum of the virtual photon, $S=(q+k)^2$ and of course $T=0$.

In the case of off-shell quark-virtual photon forward (Compton) amplitude, we have negative virtual photon square momentum ($q^2$), positive center of mass energy $S$ and null $T$, while we choose negative virtual quark square momentum ($k^2$).

In principle one should foresee a singularity in the amplitude at $T=0$. However this is not the case for helicity/angular momentum reasons. A vanishing angular momentum in the crossed ($T$) channel $q+\bar{q}\rightarrow \c^{*}+\c^{*}$ is excluded by helicity conservation and consequently the amplitude is regular at $T=0$. The amplitude is not analytic for $S>0$, as expected, indeed we are studying the absorptive part of the amplitude.

However if we complete our off-shell choice considering an unphysical negative $S$ together with the negative square momenta of the virtual scattering particles, we expect analyticity for  infrared power counting \cite{Zimmermann} reasons. Therefore the general idea is the following.
\begin{itemize}
\item We start computing the amplitudes with negative $k^2$, $q^2$ and $S$, where they are analytic. 

\item We separate the singular part when $k^2\rightarrow 0^-$. The amplitudes are expected to diverge as $\log(-k^2)$ and in fact they do. We compute the coefficient of $\log(-k^2)$ in the Mellin-Barnes expansion of each amplitude. The Mellin-Barnes expansion gives a complete $k^2$-expansion of the amplitude, however in the present analysis we disregard the contributions which are  regular in $k^2=0$.

\item The coefficient of $\log(-k^2)$ is expected to correspond to an analytic function of $S$, in the region of negative $S$ and $q^2$. By the Mellin-Barnes formula and sector decomposition we compute explicitly the analytic coefficient of $\log(-k^2)$ (as a matter of fact it is the sum of analytic contributions) and we analytically continue it in the $Re(S)>0$ region, where we find the expected branch-cut. We compute the discontinuity, which is directly related to the singular part of the trace of the partonic tensor under study.
\end{itemize}
We repeat this analysis for each graph, noticing however that, on account of  the LSZ formula, diagrams d and e have to be divided by 2.

For each spinorial amplitude  we compute the trace of the spinorial matrix multiplied by $\ksl$, since we want to sum over the helicities in the (logarithmically singular) mass-shell limit.

In momentum representation and in the Feynman gauge the box amplitude is
\beq
A^{(a)}_{\mu,\nu}(k,q)=i\,\a_s\,e^2c_F\int \frac{d^4 p}{(2\pi)^4} \,\frac{Tr[\ksl\c^{\rho}\psl\c_{\nu}(\qsl+\psl)\c_{\mu}\psl\c_{\rho}]}{(p^2)^2(q+p)^2(k-p)^2}\ ,
\label{A1}
\eeq
where $c_F$ is the quadratic Casimir of the gauge group. The box contribution to the hadronic tensor is:
\beq
W^{p,(a)}_{\mu,\nu}(k,q)=Im(A^{(a)}_{\mu,\nu}(k,q))\ .
\eeq

Let's start computing  the collinear divergent box contribution to the trace (in Lorentz indices $\mu$ and $\nu$) of the partonic tensor. The trace of the hadronic tensor gives us the radiative corrections to a linear combination \cite{Ellis:1991qj} of structure functions. The radiative corrections to an independent linear combination are given contracting the expression in Eq.(\ref{A1}) with $k^{\mu}k^{\nu}$.   We shall see that this contraction does not contain singular contributions in $k^2\rightarrow 0^{-}$.

Computing the trace we get: 
\beq
Tr[\ksl\c^{\rho}\psl\c^{\nu}(\qsl+\psl)\c_{\nu}\psl\c_{\rho}]=16\(2p\cdot k\, p\cdot q+p^2(p\cdot k-k\cdot q)\).
\eeq
At this point we pass to the Schwinger parametric form following the procedure shown in the former section and disregarding the terms proportional to $k^2$ in the numerator since these give vanishing contribution in the $k^2\rightarrow 0$ limit.

We get
\beq
A_T^{(a)}=K\,k\cdot q\int_0^{\infty}\frac{\prod_{l=1}^4d\a_l}{P_a(\a)^3}\(-i\(1+\frac{3\a_2}{P_a(\a)}\)+\frac{\a_2\,D_a(p,\a)}{P_a(\a)^2}\)e^{i\frac {D_a(p,\a)}{P_a(\a)}}
\eeq 
with $K= \a_s\,e^2c_F/\pi^2$.
Then, integrating over the scale factor $t$ and hence passing from the $\a$ to the $\b$-parameters, we get:
\beq
A_T^{(a)}=K\,k\cdot q\int\frac{d\mu(\b)}{P_a(\b)^2}\(1+\frac{2\b_2}{P_a(\b)}\)\frac{1}{D_a(p,\b)}
\label{Waq}
\eeq
with
\bea
P_a(\b)&=&\b_1+\b_2+\b_3+\b_4\\\nonumber
D_a(p,\b)&=&k^2\b_4(\b_1+\b_3)+q^2\b_2(\b_1+\b_3)+S\b_2\b_4+i\eta\ .\label{PD1}
\eea

We have shown explicitly the infinitesimal imaginary term $i\eta$, which accounts for the time-ordering in the Feynman amplitudes.
\normalsize
The amplitude is the sum of different terms corresponding to different Speer sectors. We  focus on the subset of sectors in which the  amplitude of the box has collinear singularities. These are 6 sectors corresponding to the complete cuts crossed by the quark momenta and by the (vanishing) momentum transfer. They correspond to the parametrizations:
\beq
\begin{array}{c l} 
a)&\b_1=s\quad{\rm or} \quad s\b,\ \b_2=s\, \a,\ \b_3=s\b\quad{\rm or}\quad  s,\ \b_4=1\ ;\\
b)&\b_1=1\quad{\rm or} \quad s\b,\ \b_2=s\  \a,\ \b_3=s\b\quad{\rm or} \quad 1,\ \b_4=s\ ;\\
c)&\b_1=1\quad{\rm or} \quad s,\quad \b_2=s\, \a,\quad\b_3=s\quad{\rm or} \quad 1,\  \b_4=s\b\ .\\
\end{array}
\label{abc}
\eeq
In this equation each  line corresponds to two sectors, giving the same contribution because of the symmetry of Eq.(\ref{Waq}) for the exchange of the parameters $\b_1$ and $\b_3$ . 

Once selected the interesting sectors, we consider again the expression for the amplitude given in Eq.(\ref{Waq}), which is the sum of two terms. The second term, which is proportional to $\b_2$, does not contribute to the collinear divergence, since $\b_2$ vanishes together with $D_F(p, \b)$ for $k^2= 0$.

Thus the singularity is only due to the first term in the integrand of Eq.(\ref{Waq}) and hence we must evaluate the sum of the contributions to
\beq
A_T^{(a,1)}=K\,k\cdot q\int\frac{d\mu(\b)}{P_a(\b)^2}\frac{1}{D_a(p,\b)}
\nonumber
\eeq
from the six sectors $a),\ b)\ $  and $\ c)$, that is:
{\small\bea&&
A_T^{(a,1)}=2\,K\,k\cdot q\int_0^1  { s\,ds\,d\a\,d\b\over(1+s(1+\a+\b))^2} \({1\over(1+\b)(k^2+q^2\,s\,\a)+S\,\a+i\eta}\right.\nonumber\\
 &&\left.+{1\over(1+\b\,s)(k^2+q^2\,\a)+S\,s\,\a+i\eta}
 +{1\over(1+s)(k^2\,\b+q^2\,\a)+S\,s\,\a\,\b+i\eta}\).\nn
\eea}
In order to single out the collinear divergence we apply the Mellin-Barnes formula, Eq.(\ref{MB2}), which gives:
\bea
&&A_T^{(a,1)}=-2\,K\,k\cdot q \int_{{\cal P}\[-i\infty,+i\infty\]} \frac{d\sigma}{2\pi i}(-k^2)^{-\sigma}{\Gamma(\sigma)}\Gamma(1-\sigma)\label{abo}\\&&
\int_0^1 \frac{s\,ds\,d\b\, d\a}{(1+s(1+\a+\b))^{2}}\(\frac{(1+\b)^{-\s}}{\[-\a(q^2(1+\b)s+S)-i\eta\]^{1-\s}}+\right.\nonumber\\
&&\left.\frac{(1+\b\,s)^{-\s}}{\[-\a(q^2(1+\b\,s)+S\,s)-i\eta\]^{1-\s}}+\frac{(1+s)^{-\s}}{\[-\a(q^2(1+s)+S\,s\,\b)-i\eta\]^{1-\s}}\)\ .\nonumber
\eea

Let us perform first the $\a$ integral; this has the form:
$$\int_0^1\frac{d\a}{\a}\a^{\s}\,A(\a)\ ,$$
where $A(\a)=1/\[1+s(1+\b+\a)\]^2$ is analytic and non-vanishing in the integration domain. It follows that the above integral is equal to $A(0)/\s +{\cal R}(\s)$, where ${\cal R}(\s)$ is analytic in the whole complex $\s$-plane.

The pole in $\s=0$ is the singularity considered in the former section which must be enclosed in the path encircling the $\Gamma(\s)$ poles. Thus the $\s$-integrand has a double pole in $\s=0$, while the pole would have been simple in the absence of collinear singularities.

Notice that using dimensional regularization as IR-regularization, the new pole would be in $\s=\eps$. In this case, performing the $\s$-integral one would obtain a term $-1/\eps$ from the $\Gamma$-pole in $\s=0$ and $(-k^{2})^{-\s}/\eps $ from the new pole. Then one could "go to the limit" $k^2\rightarrow 0$. Choosing, arbitrarily, $\eps$ real and negative, the second term would vanish in the limit, leaving a ${1/\eps}$ singularity, as a memory of the collinear one.

Following our strategy we  compute the $\s$-integral in Eq.(\ref{abo}),  and we select the contribution of the double pole since  the rest gives analytical contribution in $k^2\approx 0$. We get:
\bea&&
\left.A^{(a)}_T\right|_{div}=K\,\frac{2\,k\cdot q}{-q^2} \,\log(-k^2)\int_0^1 \frac{s\,ds\,d\b}{(1+s(1+\b))^{2}}\\&&
\(\frac{1}{(1+\b)s+\frac{S}{q^2}-i\eta}+\frac{1}{1+\b\,s+\frac{S}{q^2}\,s-i\eta}+\frac{1}{1+s+\frac{S}{q^2}\,s\,\b-i\eta}\)\ .\nonumber
\eea
This is an analytic function of the variable $S/q^2$ with a branch-cut in the negative real axis. We compute the discontinuity and we set
\beq
\frac{S}{q^2}=1-\frac{1}{x}\ ,
\eeq
where the new variable $0\leq x\leq 1$ is the Bjorken variable.

Setting  $z=1+s(1+\a+\b)$ we find the discontinuity:
\bea
\left.W^{p,(a)}_T\right|_{div}&=&\pi\,K\,\log(-k^2)\int_0^1 ds\int_{1+s}^{1+2s}\frac{dz}{z^{2}}\\
&&\[\d(x\,z-1)+\d(x\,z-s)+\d(-z(1-x)+1+s)\]\ .\nonumber
\eea
Notice that $-2k\cdot q/q^2=1/x$ and hence $S=q^2+2k\cdot q=q^2(1-1/x)$.
Computing the integral factor we get
\bea&&
 \frac{3x-1}{2}\theta(1-2x)\theta(3x-1)+\frac{1-x}{2}\theta(2x-1)\theta(1-x)\nn
+ x\,\theta(x)\theta(1-3x)+(1-2x)\,\theta(1-2x)\theta(3x-1)\nn
+\frac{1-3x}{2}\,\theta(x)\theta(1-3x)\ ,
\eea
where different lines correspond to different sectors and to different physical regions.

Finally the sum  of all the contributions gives:
\beq \left.W^{p,(a)}_T\right|_{div}={\a_s\,e^2c_F\over\pi}\log(-k^2)\frac{(1-x)}{2} \label{Wa}\eeq
which is the box contribution to the collinear divergent imaginary part of the amplitude  we were looking for.

Before studying the other diagrams, we still have to contract the box amplitude Eq.(\ref{A1}) with $k^{\mu}k^{\nu}$.
The numerator in momentum representation is proportional to:
\beq
Tr[\ksl\c^{\rho}\psl\ksl(\qsl+\psl)\ksl\psl\c_{\rho}]=-32(k\cdot p)^2(k\cdot p+k\cdot q)\ .\eeq
One has
\beq
k^{\mu}k^{\nu}A_{\mu\nu}^{(a)}=-2{\a_s\,e^2c_F\over\pi^2}(k\cdot q)^3\int d\mu(\b)\frac{\b_2^2(\b_1+\b_3+\b_4)}{P_a(\b)^2\,D_a(p,\b)^2}\ .
\eeq
It is very easy to verify that the amplitude is not divergent in the limit $k^2\rightarrow 0$. Indeed, in analogy with the above case we have a $\b_2^2$-proportional term and $\b_2^2$ vanishes with $D_a(p,\b)^2$ computed for $k^2= 0$.

Considering the contribution from diagram b we have to compute the first order in $\a_s$  vertex correction which is given by:
{\small\bea
\Delta^{\mu}&=i{\a_S\,e\, c_F\over 2\pi}\int &\frac{\prod_{l}^3d\b_l\d(1-\sum_{i}^3\b_i)}{(P_G(\b))^{3}}\nn\[\frac{\(\ \b_3\ \ksl-\b_2\ \qsl\ \)\c^{\mu}\(\ \qsl\ \(\b_1+\b_3\)+\ \ksl\ \b_3\)}{(k^2\,\b_1\b_3+q^2\,\b_1\b_2+S\,\b_2\b_3+i\eta)}\right.\nonumber\\
&&\left.-\c^{\mu}\,\log\(\frac{P_G(\b)^2\mu^2}{k^2\,\b_1\b_3+q^2\,\b_1\b_2+S\,\b_2\b_3}\)\].\nonumber
\label{Delta2}
\eea}
Then we can write the amplitude of  diagram b in terms of $\Delta^{\mu}$:
\beq
A^{(b)}_{\mu,\nu}(k,q)=-i\frac{e}{4\pi}\,\frac{Tr(\ksl\,\c^{\nu}\,(\ksl+\qsl)\Delta^{\mu})}{S+i\eta}\ .
\label{Apb}
\eeq
In much the same way as for the box amplitude, we first contract this tensor with the metric tensor and then with $k^{\mu}k^{\nu}$.

Contracting Eq.(\ref{Apb}) with the metric tensor, multiplying by two, since the diagrams in the figure give the same contribution, and finally disregarding the terms which are proportional to $k^2$, we have:
\bea
2A^{(b)}_{T}\approx& K\frac{2\,k\cdot q}{S+i\eta}\int \frac{d\mu(\b)}{P_b(\b)^3}&\[(\b_1+\b_3)\frac{(S-q^2)\b_3-(S+q^2)\b_2}{D_b(p,\b)}+\right.\nonumber\\
&&-\log\(\frac{P_b(\b)^2\mu^2}{D_b(p,\b)}\)\left.\]\ .
\label{tb}
\eea
with
\bea
P_b(\b)&=&\b_1+\b_2+\b_3\nonumber\\ 
D_b(p,\b)&=&k^2\b_1\b_3+q^2\b_1\b_2+S\b_2\b_3+i\eta\ .\nonumber
\eea
The  sectors contributing to the collinear singularity are those corresponding to the  parametrizations:
\bea 
&a)\,&\b_1=s,\,\b_2=s\,\a,\,\b_3=1\ ;\nonumber\\
&b)\,&\b_1=1,\,\b_2=s\,\a,\,\b_3=s\ .\nonumber
\eea
In Eq.(\ref{tb}) we have put into evidence two terms. The second one comes from the subtracted UV-divergence. It is collinear finite, since it is a parametric integral of the logarithm of a function which vanishes on the boundary of the integration domain in the $k^2\rightarrow 0^-$ limit.
The only contribution which is collinear divergent in the sectors $a)$ and $b)$, comes from the first term, that is:
\beq
2A_T^{(b,1)}=K\frac{2\,k\cdot q}{S+i\eta}\int \frac{d\mu(\b)}{P_b(\b)}(\b_1+\b_3)\frac{(S-q^2)\b_3}{D_b(p,\b)}\ ,
\eeq
the rest is finite since it is proportional to $\b_2$, which vanishes together with $D_b(p,\b)$ in much the same way as for the box. 

Summing the contributions from the two sectors, we get: 
\bea
&&2A_T^{(b,1)}=-K\frac{2\,k\cdot q}{S+i\eta}(S-q^2)\int \frac{(1+s)\,ds\, d\a}{(1+s(1+\a))^3}\nn\(\frac{1}{-k^2+\a(-q^2\,s-S)-i\eta}+\frac{s}{-k^2+\a(-q^2-S\,s)-i\eta}\)\ .\label{nuova}
\eea
We apply the Mellin-Barnes transform and we consider the collinear divergent part.\\ This is given by:
\bea
\left.2A^{(b)}_T\right|_{div}= &&K\frac{2\,k\cdot q}{S+i\eta}(1-\frac{S}{q^2})\log(-k^2)\int_0^1 \frac{ds}{(1+s)^2}\nn
\(\frac{1}{s+\frac{S}{q^2}-i\eta}+\frac{s}{1+\frac{S}{q^2}\,s-i\eta}\)\ .\label{xyz}\eea
Again this is an analytic function of the variable $S/q^2$ with a branch-cut on the negative real axis and a pole in $-i\eta$. 

The novelty of the vertex correction contribution lies in the presence of the pole superimposed on the branch cut. Thus computing the absorptive part we should take into account the branch-cut discontinuity and the contribution proportional to $-i\pi\delta(S)$ coming from the pole. This corresponds to a single parton final state which is also present in the collinear-divergent contributions from diagrams d and e.

However the term $i\pi\delta(S)$ multiplying the vertex  
correction in diagrams b and c is ill-defined, since we see from  
Eq.(\ref{xyz}) that the coefficient of the Dirac delta contains the  
ill-defined integral $\int_0^1ds/[s(1+s)^2]\ .$ Furthermore the integral factor in Eq.(\ref{xyz}) has a branch cut whose discontinuity diverges when $S$ vanishes.

  A general remark is here in order. Even if the Fourier transformed  
Feynman amplitude corresponding to a given diagram is the product of  
the contributions from the parts of the diagram, this product  
structure is not suitable for spectral analyses. Indeed for this  
purpose one has to introduce a spectral representation for the whole  
amplitude and compute, e.g. its imaginary part, on the basis of this  
representation.

In the light of this comment it is easy to verify that the two above mentioned inconsistencies compensate each other. Indeed let us look more carefully at Eq.(\ref{xyz}). It can be written as follows:
\bea
\left.2A^{(b)}_T\right|_{div}&=&K\,\frac{\log(-k^2)}{x}\,\[\int_0^1 \frac{ds}{s\,\(1+s\)^2}\(\frac{1}{s+\frac{S}{q^2}-i\eta}-\frac{1}{\frac{S}{q^2}-i\eta}\)+\right.\nonumber\\
&&-\frac{1}{S/q^2-i\eta}\int_0^1 \frac{s\,ds}{\(1+s\)^2}\frac{1}{\frac{S}{q^2}\,s-i\eta}+\int_0^1\frac{ds}{\(1+s\)^2}\cdot\nonumber\\
&&\left. \(\frac{1}{s+\frac{S}{q^2}-i\eta}-\frac{s}{1+s\,\frac{S}{q^2}-i\eta}\)\right]\ .
\label{Apb3}
\eea
Now, considering the imaginary part and setting 
\beq
\frac{S}{q^2}=1-\frac{1}{x},
\nonumber\eeq
we have:
\bea
\left.2W^{p,(b)}_T\right|_{div}&=&K\,\pi\,\log(-k^2)\,x\,\[\int_{1/2}^1 \frac{dt}{1-t}t\(\delta(x-t)-\delta(x-1)\)\right.\nonumber\\ && \left.
+\int_{1/2}^1 dt\({1\over1-x}\delta(x-(1-t))+\delta(x-t)\)\right.\nonumber\\ &&\left.
-\delta(x-1)\(\log(2)-{1\over2}\)\],\nonumber
\eea
where we have  changed the integration variable $t=1/(1+s)$.

One might wonder  if this is a distribution in $0\leq x\leq 1$. To check this point, we multiply $2W^{p,(b)\,\nu}_{\nu}\left.\right|_{div}$ by a $C_{\[0,1\]}^{\infty}$ function $\phi(x)$ and integrate over $x$. \\
We get:
\bea
\int_0^1 dx\phi(x)\left.2W^{p,(b)}_T\right|_{div}&&=K\,\pi\,\log(-k^2)\[\int_{1/2}^1 dx\,\frac{x}{1-x}\(\phi(x)-\phi(1)\)\right.\nonumber\\
&&+\left.\int_0^{1/2}dx\,\frac{x}{1-x}\phi(x) -(\log2-\frac{1}{2})\phi(1)\]=\nonumber\\
&&K\,\pi\,\log(-k^2)\[\int_0^1 dx\frac{x\phi(x)-\phi(1)}{1-x}+\phi(1)\]\ 
.\label{Wpb}
\eea
Defining as in the literature \cite{Ellis:1991qj} the distribution $\frac{1}{(1-x)_+}$ as
\beq
\int_0^1 dx\frac{f(x)}{(1-x)_+}=\int_0^1dx\frac{f(x)-f(1)}{1-x}\ ,
\eeq
we identify Eq.(\ref{Wpb}) with:
\beq
\left.2W^{p,(b)}_T\right|_{div}={\a_s\,e^2c_F\over\pi}\log(-k^2)\(\frac{x}{(1-x)_+}+\delta(1-x)\)\ .
\eeq
We sum this result with Eq.(\ref{Wa}) getting
\beq
\left.(W^{p,(a)}_T+2W^{p,(b)}_T)\right|_{div}=\frac{\a_s\,e^2\,c_F}{2\,\pi}\,\log(-k^2)\(\frac{1+x^2}{(1-x)_+}+2\,\delta(1-x)\)\ .
\eeq
Now one can compute the splitting function $P(x)$\cite{Ellis:1991qj}
\beq
P(x)=c_F\(\frac{1+x^2}{(1-x)_+}+2\,\delta(1-x)+C\,\delta(1-x)\)\ ,
\eeq
where $C$ is the coefficient of the contribution from the diagrams d and e. Notice that diagram f does not contribute to the collinear-divergent part of $g^{\a,\b}W_{\a,\b}$ since it is proportional to the tree diagram term multiplied by $\log(S)$.

Considering that, on account of the LSZ formula, diagrams d and e have to be divided by 2, one has 
\beq
\left.(W^{p,(d)}_T+W^{p,(e)}_T)\right|_{div}=-\frac{1}{4}\frac{\a_S\,e^2 c_F}{\pi}\,\log(-k^2) \,x\, \delta(1-x)\ .
\eeq
Hence $C=-\frac{1}{2}$ and we get the known value of the splitting function:
\beq
P(x)=c_F\(\frac{1+x^2}{(1-x)_+}+\frac{3}{2}\delta(1-x)\)\ .
\eeq
We have still to contract $2A^{(b)}_{\mu,\nu}$ with $k^{\mu}k^{\nu}$. One easily sees that the result vanishes in the collinear limit. 

A final remark is here in order concerning the gauge independence of our results~\cite{Dorn:2008dz}.
This can be proved extending the basic argument for the
gauge independence of the charged particle mass shell amplitudes in massive QED
 \cite{Lowenstein:1972pr}. Indeed in any gauge theory there
is a general connection between the gauge independence of the
"physical" amplitudes and Ward-Slavnov-Taylor identities \cite{Becchi:1974md}. In the case of amplitudes involving only local
physical (BRS invariant) operators one finds directly that their
partial derivative with respect to any gauge fixing parameter
vanishes. The case of amplitudes involving charged fields, which are
not BRS invariant, is more difficult, indeed one finds that the same
derivatives are less singular on the mass-shell than the original
amplitudes; poles are replaced by branch cuts. The idea which can be
verified in a general situation is that in mass-less theories the same
thing happens since any gauge parameter derivative of a collinear
divergent amplitude is less singular, or even null in the mass-shell
limit. This is fairly clear in our DIS example, indeed, using
Slavnov-Taylor identities, one shows that the Feynman-gauge parameter
derivative of the non-amputated forward scattering amplitude is
proportional to the box diagram amplitude in which the gluon
propagator has been replaced by a scalar propagator. The result is
apparently proportional to the box amplitude (there is a $-2$ factor),
therefore it is proportional to $\log (-k^2)$. However in order to get the
derivative of the forward scattering amplitude one has to amputate the
amplitude multiplying it by $k^2$. This kills the logarithmic collinear
divergence proving that the gauge derivative of the amputated amplitude vanishes in the
mass-shell limit.

\section{Conclusions}
We have seen  how one can study collinear
divergences in DIS cross sections by considering the quark mass-shell
limit of Feynman amplitudes and exploiting the simplifying power of
the Speer-Smirnov sector decomposition of the parametric integral
expressions for the amplitudes together with  the Mellin-Barnes transform.

In particular, we have shown that the off-shell regularization is a
natural tool for the analysis of collinear divergences in inclusive
cross sections, and that the use of this regularization is strongly
simplified by the Speer-Smirnov sector decomposition. The recourse to
Mellin-Barnes formula is the obvious final tool for the computation of the
singular parts of the amplitudes.

The suggestion of initial state off-shell regularization of inclusive
cross sections as a reasonable "physical option", as an alternative 
to the widely adopted dimensional scheme,
is one of the results of the present paper. 
However the principal goal of this paper was to identify a general
method of analysis of mass singularities in the Feynman amplitudes of
massless field theories, which could be automatically applied to
multi-loop and multi-external-vertex diagrams. For this reason we have
described with great care the role of different kinds of cuts in the
construction of the parametric form of a Feynman integral. These cuts
are sets of lines in Feynman diagrams whose deletion either breaks the
original diagram in two parts, or transforms it into a tree diagram,
or else does both things together. We have also analyzed
the role of the cuts in the identification of the Speer-Smirnov
s-families and sectors. The idea is
that the identification of sectors contributing to the IR-singular
parts associated with the vanishing of kinetic invariants is a
central step in the study of these singularities, and that the method
could be extended to multi-loop and multi-external-vertex diagrams
developing suitable software tools. As a matter of fact, in the last
year an example of such tools (Fiesta 2 ~\cite{Smirnov:2009pb}) was
developed by Smirnov and Tentyukov. However, to our knowledge, the
method has not yet been used in the explicit analysis of collinear
divergences.

A systematic use of the techniques presented in this paper may in principle
prove useful in different contexts of phenomenological
interest. High-order perturbative calculations
in QCD, for example, may take advantage of a systematic separation of
singular contributions.
From a different point of view, an efficient way of identifying IR divergent
terms to all orders in perturbation theory can be useful when one is faced
with the problem of resumming the whole perturbative series in
special kinematic regimes, where powers of large logarithms of ratios of
invariants spoil the reliability of fixed-order calculations.

A  point that we have not discussed with sufficient completeness is the
effect of the UV subtractions parts on the IR analysis. 
 A further study on this point  is crucial if one tries to work with high-order
diagrams. One should compare Speer-Smirnov
s-families with Hepp-Zimmermann forests and show that the UV subtraction should be limited to the sectors whose corresponding s-families contain  one-particle-irreducible UV divergent components. 

In conclusion, being clear that
Speer's sector decomposition is useful in the singularity analysis,
while it does not seem particularly suited for regular diagram
calculations for which other sector decompositions have been developed
together with a suitable software \cite{pilipp} \cite{binoth} \cite{Carter:2010hi} , we believe that our work has been
convincing enough concerning the advantages of the Speer-Smirnov
sector decomposition combined with the Mellin-Barnes formula in the
IR- singularity analysis, also in view of the study of multi-loop
multi-leg amplitudes. This point is also stressed by A.V. Smirnov,
V.A. Smirnov and M. Tentyukov in some recent works
\cite{Smirnov:2008py}~\cite{Smirnov:2009pb}.
One should however notice that, when the number of legs increases together with that of the corresponding kinetic invariants,  the analytic continuation foreseen in our strategy might become prohibitive. It is clear that the discussed example and  the Drell-Yan processes, in which the analytic continuation concerns a single variable, the total energy,  correspond to particularly lucky situations.

\section*{Acknowledgments} We are grateful to G.Ridolfi for constant encouragement and help in the preparation of this paper.


\end{document}